\documentclass[a4paper,11pt]{amsart}

\usepackage[colorlinks=true, linkcolor=red, linktoc=page]{hyperref}
\usepackage{amssymb}
\usepackage{enumitem}
\usepackage{url}
\usepackage{xcolor}
\usepackage{graphicx}
\usepackage{geometry}

\usepackage{algorithm}
\usepackage{algorithmic}

\usepackage[T1]{fontenc}
\usepackage[english]{babel}
\usepackage{times}
\usepackage{ucs}

\usepackage{dsfont}
\usepackage[mathscr]{euscript}
\usepackage{stmaryrd}

\makeatletter

\def\section{\@startsection{section}{1}%
\z@{.7\linespacing\@plus\linespacing}{.5\linespacing}%
{\normalfont\bfseries\centering}}

\def\@settitle{\begin{center}%
  \baselineskip14\p@\relax
    \bfseries
    \LARGE\@title
  \end{center}%
}

\def\@setauthors{%
  \begingroup
  \trivlist
  \centering\footnotesize \@topsep30\p@\relax
  \advance\@topsep by -\baselineskip
  \item\relax
  \andify\authors
  \def\\{\protect\linebreak}%
 {\Large\authors}%
  \endtrivlist
  \endgroup
}

\def\maketitle{\par
  \@topnum\z@ 
  \@setcopyright
  \thispagestyle{firstpage}
  \ifx\@empty\shortauthors \let\shortauthors\shorttitle
  \else \andify\shortauthors
  \fi
  \@maketitle@hook
  \begingroup
  \@maketitle
  \toks@\@xp{\shortauthors}\@temptokena\@xp{\shorttitle}%
  \toks4{\def\\{ \ignorespaces}}
  \edef\@tempa{%
    \@nx\markboth{\the\toks4
      \@nx{\the\toks@}}{\the\@temptokena}}%
  \@tempa
  \endgroup
  \c@footnote\z@
  \def\do##1{\let##1\relax}%
  \do\maketitle \do\@maketitle \do\title \do\@xtitle \do\@title
  \do\author \do\@xauthor \do\address \do\@xaddress
  \do\email \do\@xemail \do\curraddr \do\@xcurraddr
  \do\commby \do\@commby
  \do\dedicatory \do\@dedicatory \do\thanks \do\thankses
  \do\keywords \do\@keywords \do\subjclass \do\@subjclass
}
\makeatother

\newtheorem{defi}{Definition}[section]

\newtheorem{rk}[defi]{Remark}

\newcommand{\Exp}{\mathbb{E}}

\renewcommand{\Pr}{\mathbb{P}}

\newcommand{\eps}{\varepsilon}

\title{Survival probability of structures under fatigue: a data-based approach}

\author[F.-B. Cartiaux, F. Legoll, A. Libal and J. Reygner]{Fran\c cois-Baptiste Cartiaux$^1$, Frédéric Legoll$^{2,3}$,\\ Alex Libal$^{2,4}$ and Julien Reygner$^4$\\ \vskip 0cm
{\footnotesize $^1$ OSMOS Group, Puteaux, France}\\
{\footnotesize $^2$ Navier, École des Ponts, Univ Gustave Eiffel, CNRS, Marne-La-Vall\'ee, France}\\
{\footnotesize $^3$ MATHERIALS project-team, Inria, Paris, France}\\
{\footnotesize $^4$ CERMICS, École des Ponts, Marne-La-Vall\'ee, France}\\ \vskip 0cm
{\footnotesize \tt cartiaux@osmos-group.com, \{frederic.legoll,ales.libal,julien.reygner\}@enpc.fr}}

\date{\today}
\thanks{}

\begin{document}

\begin{abstract}
  This article addresses the probabilistic nature of fatigue life in structures subjected to cyclic loading with variable amplitude. Drawing on the formalisation of Miner's cumulative damage rule that we introduced in the recent article~\cite{Car23}, we apply our methodology to estimate the survival probability of an industrial structure using experimental data. The study considers both the randomness in the initial state of the structure and in the amplitude of loading cycles. The results indicate that the variability of loading cycles can be captured through the concept of deterministic equivalent damage, providing a computationally efficient method for assessing the fatigue life of the structure. Furthermore, the article highlights that the usual combination of Miner's rule and of the weakest link principle systematically overestimates the structure's fatigue life. On the case study that we consider, this overestimation reaches a multiplicative factor of more than two. We then describe how the probabilistic framework that we have introduced offers a remedy to this overestimation.
\end{abstract}

\maketitle

\baselineskip=16pt


\section{Introduction}

The fatigue life of specimens or structures subjected to deterministic loading is commonly observed to display a significant statistical dispersion. This has led to the understanding of fatigue as an intrinsically probabilistic phenomenon, owed to the presence of randomly distributed microscopic flaws, such as cracks, in the material~\cite{CasFer09}. Many works have been devoted to the construction of statistical models for fatigue testing experiments~\cite{BirSau68,Sau70,Cas06,Thi08,Fou14,Bar19,Mir20,Aen23}. On the other hand, the standard stress-based approach to the fatigue life of specimens or structures subjected to cyclic loading with variable amplitude~\cite[Chapter~9]{Dow13}, based for instance on Miner's cumulative damage rule~\cite{Pal24,Min45}, consists in computing deterministic quantities which are taken as representative values of the lifetime of the specimen or the structure. Our recent article~\cite{Car23} introduces a probabilistic formalisation of Miner's rule, which gives an interpretation of these deterministic quantities in terms of the true, random lifetime of the specimen or the structure.

In the present article, this framework is applied to estimate the survival probability of an actual industrial structure, on the basis of experimental data measured by sensors installed by OSMOS Group on the structure. In this context, both the initial state of the structure and the amplitude of loading cycles are assumed to be random. A particularly interesting feature of this case study is that the sensors do not measure the amplitude of the loading cycles per se. As explained below, they rather record deformation cycles (over a given period of time) at various locations of the structure. Some data may be correlated (because they correspond to a single physical event), and we will not assume to have at our disposal the correlation structure. 

In short, the conclusion of this work is twofold. On the methodological side, our first conclusion is that the variability of loading cycles can be summarised by a notion of deterministic equivalent damage, which provides a fast and easy-to-implement method to assess the fatigue life of the structure. On the practical side, our second conclusion is that the usual combination of Miner's rule and of the weakest link principle leads to a systematic overestimation of the structure's fatigue life. In our case study, this overestimation reaches a multiplicative factor of more than two. As shall be explained below, the use of the probabilistic formalism that we introduced in~\cite{Car23} allows to avoid this overestimation.

This article is organised as follows. The theoretical ingredients of our work are presented in Section~\ref{s:miner}. They are based on our previous study~\cite{Car23} and also contain novel developments, including the introduction of the notion of deterministic equivalent damage (see~\eqref{eq:Dkeq} below). The case study we consider is presented in Section~\ref{s:case}, and numerical results are reported in Section~\ref{s:curves}. The conclusions of our investigations are summarised in Section~\ref{s:conclusion}.

\section{Probabilistic formalisation of the stress-based approach}\label{s:miner}

The building bricks of the stress-based approach to fatigue are S-N curves and Miner's cumulative damage rule, which are respectively introduced in Section~\ref{ss:sn} and~\ref{ss:miner}. The application of this approach to structural reliability when only the randomness on the initial state of the structure is taken into account is next described in Section~\ref{ss:struct}. The randomness on the severity of loading cycles is then addressed in Section~\ref{ss:rand-load}, where the notion of deterministic equivalent damage is introduced.

\subsection{S-N curves}\label{ss:sn}

In fatigue testing experiments, standardised specimen are subjected to cyclic loading with constant amplitude. The overall stress amplitude of each cycle is summarised by a scalar positive quantity $S$, called the \emph{severity} of the cycle and expressed in $\mathrm{MPa}$. For a given severity $S$ and a reference probability $p \in (0,1)$, the \emph{S-N curve} indicates the number of cycles $N_p(S)$ at which a proportion $p$ of specimens has failed~\cite{Woh70}. 

In probabilistic terms, the Number of Cycles to Failure (NCF) of a specimen subjected to cyclic loading with constant severity $S$ can be modelled as a random variable $N(\omega,S)$, the law of which depends on $S$. The abstract parameter $\omega$ encodes the variability of the initial state of the specimen, namely the distribution of microscopic flaws. It is not given a precise physical meaning: its statistical distribution is only measured through the observation of independent and identically distributed (i.i.d.) realisations of the random variable $N(\omega,S)$. In this setting, the deterministic quantity $N_p(S)$ provided by the S-N curve is the \emph{quantile} of order $p$ of $N(\omega,S)$: we have $\Pr\left(N(\omega,S) \leq N_p(S) \right) = p$.

\subsection{Miner's rule}\label{ss:miner}

Miner's cumulative damage rule allows to compute the NCF of a specimen subjected to cyclic loading with \emph{variable} amplitude, described by a severity sequence $\mathbf{S} = (S_n)_{n \geq 1}$ where $S_n$ is the severity of the $n$-th cycle. In the present section, this sequence is considered deterministic, so the randomness of the specimen's lifetime only originates in the variability of its initial state. The cumulative damage after $n$ cycles is defined by
\begin{equation}\label{eq:Dpn}
    D_{p,n}(\mathbf{S}_n) = \sum_{i=1}^n \frac{1}{N_p(S_i)},
\end{equation}
where $\mathbf{S}_n = (S_i)_{1 \leq i \leq n}$ and, at the $i$-th cycle, $N_p(S_i)$ is given by the S-N curve. Miner's NCF of the specimen is then defined by
\begin{equation}\label{eq:Np}
    N_p(\mathbf{S}) = \inf \left\{ n \geq 1: D_{p,n}(\mathbf{S}_n) \geq 1 \right\}.
\end{equation}
This deterministic quantity is taken as the representative value of the random lifetime of the specimen (expressed in number of loading cycles). The precise probabilistic interpretation of this quantity was given in~\cite{Car23}: when the severity sequence $\mathbf{S}$ is constant and equal to $S$, it follows from the construction of S-N curves that $N_p(\mathbf{S}) = N_p(S)$, that is to say that Miner's NCF is the quantile of order $p$ of the specimen's random NCF $N(\omega,S)$. The first main result of~\cite{Car23} is the fact that, under mild assumptions, this statement remains valid for an \emph{arbitrary} severity sequence $\mathbf{S}$: in other words, the NCF $N_p(\mathbf{S})$ computed by the deterministic Miner rule is the quantile of order $p$ of the specimen's random NCF $N(\omega,\mathbf{S})$ associated with the severity sequence $\mathbf{S} = (S_n)_{n \geq 1}$.

\subsection{Structure fatigue life assessment}\label{ss:struct}

In the assessment of the fatigue life of a structure, a common approach, dubbed \emph{weakest link principle}~\cite{Wei51}, consists in dividing the structure into $K \geq 1$ elementary volumes, which are considered as statistically independent specimens for which S-N curves are available. The NCF $N_p^k(\mathbf{S}^k)$ of each elementary volume $k$ (with $1 \leq k \leq K$), defined by~\eqref{eq:Np}, is then computed by Miner's rule~\eqref{eq:Dpn}, assuming that the severity sequence $\mathbf{S}^k$ to which this elementary volume is subjected can be measured. Note also that the S-N curve used for the elementary volume $k$ may be different from one volume to the next, thus the dependency upon $k$ of the function $\mathbf{S} \mapsto N_p^k(\mathbf{S})$. 

Taking the NCF $N_p^k(\mathbf{S}^k)$ of each elementary volume $k$ as a deterministic representative value of the random lifetime of the volume, the weakest link principle leads to defining the structure's NCF as the minimal NCF among elementary volumes, that is to say that one sets
\begin{equation}\label{eq:Npstruct}
    N^\mathrm{min}_p\left((\mathbf{S}^k)_{1 \leq k \leq K}\right) = \min_{1 \leq k \leq K} N_p^k(\mathbf{S}^k)
\end{equation}
and takes this deterministic quantity as a representative value of the random lifetime of the structure.

A corollary of the probabilistic interpretation of Miner's rule recalled above is that this procedure \emph{systematically overestimates} the fatigue life of the structure. Indeed, applying the weakest link principle to the \emph{random} NCF of the structure leads to defining the latter as
\begin{equation}\label{eq:Nomega}
    N\left(\omega,(\mathbf{S}^k)_{1 \leq k \leq K}\right) = \min_{1 \leq k \leq K} N^k(\omega,\mathbf{S}^k),
\end{equation}
in the sense that, for a given initial state $\omega$ and severity sequences $(\mathbf{S}^k)_{1 \leq k \leq K}$, the structure fails as soon as one of the elementary volumes fails. It then follows from the elementary but important remark that {\em the quantile of the minimum among random variables is always smaller than the minimum of the quantiles of these variables} that the actual quantile of order $p$ of $N\left(\omega,(\mathbf{S}^k)_{1 \leq k \leq K}\right)$, which we denote by $N_p\left((\mathbf{S}^k)_{1 \leq k \leq K}\right)$, is smaller than the quantity $N^\mathrm{min}_p\left((\mathbf{S}^k)_{1 \leq k \leq K}\right)$ defined in~\eqref{eq:Npstruct}. 

To quantify this statement, let us recall that the elementary volumes are disjoint and therefore considered as statistically independent, since randomness is assumed to originate in microscopic flaws. This implies that the random variables $(N^k(\omega,\mathbf{S}^k))_{1 \leq k \leq K}$ are statistically independent. We thus have that, for any $n \geq 1$, the probability that the random NCF of the structure be at least $n$ writes in terms of the elementary volumes' NCFs as follows:
\begin{equation}\label{eq:PN}
    \Pr\left(N\left(\omega,(\mathbf{S}^k)_{1 \leq k \leq K}\right) > n\right) = \prod_{k=1}^K \Pr\left(N^k(\omega,\mathbf{S}^k)>n\right).
\end{equation}
This identity allows to compute survival probability curves for the structure as a function of the number of cycles. As a by-product, it also allows to compute the quantile of order $p$ of $N\left(\omega,(\mathbf{S}^k)_{1 \leq k \leq K}\right)$, which is the deterministic representative value of the random lifetime of the structure which is consistent with Miner's rule.

In order to make this computation tractable, a common practice consists in assuming that the random variables $(N^k(\omega,\mathbf{S}^k))_{1 \leq k \leq K}$ follow a Weibull distribution, with \emph{modulus} $m_\mathrm{W}>0$ (see~\cite{Wei51,CasFer09}). In this case, the survival probability of each elementary volume can be deduced from its cumulative damage, defined by
\begin{equation}\label{eq:Dkpn}
    D^k_{p,n}(\mathbf{S}^k_n) = \sum_{i=1}^n \frac{1}{N^k_p(S^k_i)},
\end{equation}
through the identity (see~\cite[Eq.~(8)]{Car23})
\begin{equation}\label{eq:PNk-weibull}
    \Pr\left(N^k(\omega,\mathbf{S}^k)>n\right) = \exp\left(\log(1-p) \, \left[ D^k_{p,n}(\mathbf{S}^k_n) \right]^{m_\mathrm{W}} \right).
\end{equation}
As a consequence, in view of~\eqref{eq:PN} and~\eqref{eq:PNk-weibull}, the survival probability of the structure is given by the formula
\begin{equation}\label{eq:surv-struct-P}
    \Pr\left(N\left(\omega,(\mathbf{S}^k)_{1 \leq k \leq K}\right) > n\right) = \exp\left(\log(1-p) \, \sum_{k=1}^K \left[ D^k_{p,n}(\mathbf{S}^k_n) \right]^{m_\mathrm{W}} \right),
\end{equation}
which shows that the quantile $N_p\left((\mathbf{S}^k)_{1 \leq k \leq K}\right)$ of order $p$ of the random NCF of the structure is the number of cycles $n$ which satisfies
\begin{equation}\label{eq:barN-intermediaire}
    \sum_{k=1}^K \left[ D^k_{p,n}(\mathbf{S}^k_n) \right]^{m_\mathrm{W}} = 1.
\end{equation}

\subsection{Randomness of loading and deterministic equivalent damage}\label{ss:rand-load}

In this section, we now assume that the severity sequence $(\mathbf{S}^k)_{1 \leq k \leq K}$ to which each elementary volume of the structure is subjected is random, and independent from the initial state of the structure. We use the notation $\Pr^*$ and $\Exp^*$ to refer to probability and expectation taking this randomness into account. The formula~\eqref{eq:surv-struct-P} for the survival probability of the structure then becomes
\begin{equation}\label{eq:surv-struct-P*}
    \Pr^*\left(N\left(\omega,(\mathbf{S}^k)_{1 \leq k \leq K}\right) > n\right) = \Exp^*\left[\exp\left(\log(1-p) \, \sum_{k=1}^K \left[ D^k_{p,n}(\mathbf{S}^k_n) \right]^{m_\mathrm{W}} \right) \right].
\end{equation}
Since, at each cycle $n \geq 1$, the severity of the loading undergone by each elementary volume originates in a single physical event, the random variables $S^k_n$, for $1 \leq k \leq K$, are likely to be non-trivially correlated and therefore the expectation in the right-hand side of~\eqref{eq:surv-struct-P*} cannot, in general, be computed analytically. In principle, it must thus be evaluated by the Monte Carlo method.

However, if the vectors $(S^k_n)_{1 \leq k \leq K}$, for any $n \geq 1$, are assumed to be independent and identically distributed copies of some random vector $(S^k)_{1 \leq k \leq K}$, then, for any $k \in \{1, \ldots, K\}$, the Law of Large Numbers applied on~\eqref{eq:Dkpn} yields the first-order asymptotics
\begin{equation}\label{eq:lln}
    D^k_{p,n}(\mathbf{S}^k_n) = \sum_{i=1}^n \frac{1}{N^k_p(S_i^k)} \sim n \, \Exp^*\left[\frac{1}{N^k_p(S^k)}\right] \qquad \text{when $n \to +\infty$}.
\end{equation}
Thus, defining the \emph{deterministic equivalent damage} $D^k_{p,\mathrm{eq}}$ by the identity
\begin{equation}\label{eq:Dkeq}
    D^k_{p,\mathrm{eq}} = \Exp^*\left[\frac{1}{N^k_p(S^k)}\right],
\end{equation}
and using the approximation~\eqref{eq:lln} in~\eqref{eq:surv-struct-P*}, the formula for the survival probability of the structure becomes
\begin{equation}\label{eq:surv-struct-P*-eq}
    \Pr^*\left(N\left(\omega,(\mathbf{S}^k)_{1 \leq k \leq K}\right) > n\right) = \exp\left(\log(1-p) \, \sum_{k=1}^K \left(n \, D^k_{p,\mathrm{eq}}\right)^{m_\mathrm{W}} \right).
\end{equation}

\begin{rk}\label{rk:Skeq}
    The notion of deterministic equivalent damage is naturally associated with the notion of \emph{deterministic equivalent severity} $S^k_{p,\mathrm{eq}}$, which can be defined by the identity
    \begin{equation}\label{eq:Skeq}
        \frac{1}{N^k_p(S^k_{p,\mathrm{eq}})} = D^k_{p,\mathrm{eq}}, \quad \text{where we recall that} \quad D^k_{p,\mathrm{eq}} = \Exp^*\left[\frac{1}{N^k_p(S^k)}\right].
    \end{equation}
    It corresponds to the deterministic severity which produces, in average, the same cumulative damage as the random severity $S^k$.
    We also refer to the discussion in~\cite[Section~5.4]{Car23}. Interestingly, $S^k_{p,\mathrm{eq}}$ is in general not equal to the mean value of the severity sequence. Therefore, while deterministic, this quantity actually encodes the variability (i.e. some features of the probability distribution) of loading cycles.
\end{rk}

Once the deterministic equivalent damage is estimated, the computation of survival probability curves becomes straightforward (it just amounts to evaluating~\eqref{eq:surv-struct-P*-eq}) and no longer requires Monte Carlo simulations. Moreover, with the same argument as for the derivation of~\eqref{eq:barN-intermediaire}, we deduce from~\eqref{eq:surv-struct-P*-eq} that the quantile $N^*_p$ of order $p$ of the random NCF of the structure, taking into account both randomness on the initial state of the structure and on the severity of loading cycles, is the number of cycles $n$ such that
\begin{equation*}
    \sum_{k=1}^K \left(n \, D^k_{p,\mathrm{eq}}\right)^{m_\mathrm{W}} = 1,
\end{equation*}
that is to say that
\begin{equation}\label{eq:barN*}
    \frac{1}{(N^*_p)^{m_\mathrm{W}}} = \sum_{k=1}^K \left(D^k_{p,\mathrm{eq}}\right)^{m_\mathrm{W}}.
\end{equation}
With the notation introduced in Remark~\ref{rk:Skeq}, this identity also rewrites
\begin{equation*}
    \frac{1}{(N^*_p)^{m_\mathrm{W}}} = \sum_{k=1}^K \frac{1}{\left[ N^k_p(S^k_{p,\mathrm{eq}}) \right]^{m_\mathrm{W}}}.
\end{equation*}

\section{Presentation of the case study}\label{s:case}

\subsection{Context and database}

The structure under monitoring is a mobile crane in an industrial site, which is used every day. It is endowed with $K=8$ sensors, whose names are listed in Table~\ref{tab:sensors}. Each sensor records the dimensionless magnitude $\eps$ of longitudinal deformation (along the sensor) at each loading cycle. The crane is thus assumed to be divided into $K$ elementary volumes, hereafter referred to as \emph{zones}, each of which being monitored by one sensor.

The sensors named OS1 to OS4 are located at the ends of the main beams of the mobile crane, on the upper flange, where the steel is subjected to longitudinal tension at the bolted junction with the supports. The sensors named OS5 and OS6 are located on secondary transverse supports for a technical cabin aside one of the beams: their behavior is not strongly linked to the load on the crane, unlike all other six sensors, and indeed the fatigue issue does not appear stringent in their case. The sensors named OS7 and OS8 are located at the middle of the span length, under the lower flange of the main beams, where the loads induce significant longitudinal tension in the steel.

The measurements are available for $N=912$ days. The database collects, for each day and each sensor, the magnitude $\eps$ of each loading cycle detected and recorded by the sensor. The cycle magnitudes are obtained from the raw strain measurements by a classical Rainflow counting method.

\begin{table}[htpb]
    \centering
    \begin{tabular}{cl}
      $k$ & Name of the sensor\\
      \hline
      $1$ & \texttt{OS1-Back-Right}\\
      $2$ & \texttt{OS2-Front-Right}\\
      $3$ & \texttt{OS3-Back-Left}\\
      $4$ & \texttt{OS4-Front-Left}\\
      $5$ & \texttt{OS5-Support-Right}\\
      $6$ & \texttt{OS6-Support-Left}\\
      $7$ & \texttt{OS7-Front-Mid}\\
      $8$ & \texttt{OS8-Back-Mid} \\ \\
    \end{tabular}
    \caption{Names of the $8$ sensors associated with each elementary volume.}
    \label{tab:sensors}
\end{table}

\begin{rk}
    In the database, values are missing for the sensor \emph{\texttt{OS2-Front-Right}}. In order not to neglect the risk of failure of the associated zone, we simply duplicate the data from the sensor \emph{\texttt{OS1-Back-Right}}, which is symmetrically positioned on the crane and therefore can be considered to virtually monitor similar deformations.
\end{rk}

\subsection{Severity of a loading cycle}

The severity $S^k$ of a cycle undergone by the $k$-th zone is related to the deformation $\eps$ measured by the $k$-th sensor through the identity
\begin{equation}\label{eq:S-eps}
    S^k = C^k_\mathrm{s} \, E^k \, \eps,
\end{equation}
where $E^k$ is the Young modulus (in $\mathrm{MPa}$) of the material and $C_\mathrm{s}^k \geq 1$ is a dimensionless safety factor. Both $E^k$ and $C_\mathrm{s}^k$ may depend on the considered zone. In our experiments, we shall take $E^k = 210 \times 10^3\, \text{MPa}$ for all $k$, and either $C_\mathrm{s}^k = C_\mathrm{s} = 1$ for all $k$ or $C_\mathrm{s}^k = C_\mathrm{s} = 3$ for all $k$.

\subsection{Computation of daily cumulative damage}

\begin{figure}[htpb]
  \centering
  \includegraphics[width=.8\textwidth]{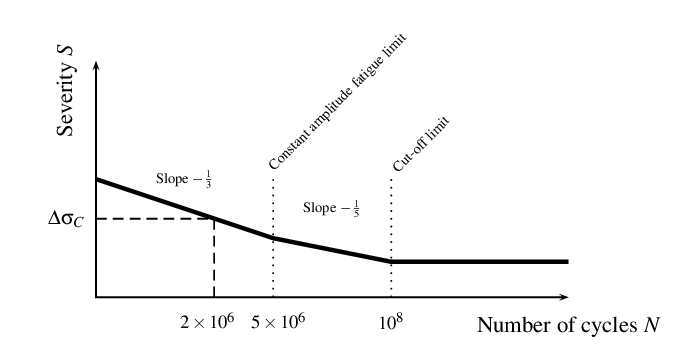}
  \caption{S-N curve used for each sensor. In loglog coordinates, it is piecewise linear, with the slopes $-1/3$ and $-1/5$ prescribed by Part~1-9 of Eurocode~3. The curve is then entirely characterised by the \emph{detail category} $\Delta \sigma_C$, which is the severity for which the NCF assumes the value $2 \times 10^6$.}
  \label{fig:sn}
\end{figure}

The damage $1/N^k_p(S^k)$ associated (at the $k$-th zone) with a cycle with severity $S^k$ is obtained by the S-N curve of Figure~\ref{fig:sn}, which is taken from Part~1-9 of Eurocode~3 and is given for a reference probability $p=0.05$. A curve is characterised by its \emph{detail category} $\Delta \sigma^k_C$, which is the severity for which $N^k_p(\Delta \sigma^k_C) = 2 \times 10^6$. The detail category may depend on the considered zone. In our experiments, we shall take either $\Delta \sigma^k_C = \Delta \sigma_C = 36\, \mathrm{MPa}$ for all $k$ or $\Delta \sigma^k_C = \Delta \sigma_C = 80\, \mathrm{MPa}$ for all $k$.

For each day $i$, the daily cumulative damage for the sensor $k$ is defined by
\begin{equation}\label{eq:dkpi}
  d^k_{p,i} = \sum_n \frac{1}{N^k_p(S^k_n)},
\end{equation}
where the sum runs over all cycles $n$ measured by the sensor $k$ on the day $i$. The sample of daily cumulative damages $\{ (d^k_{p,i})_{1 \leq k \leq K}, \ 1 \leq i \leq N\}$ is considered as a set of $N$ i.i.d. realisations of some $K$-dimensional vector $(D^k_p)_{1 \leq k \leq K}$, with dimensionless coordinates. The marginal histograms of each of the $K$ coordinates and the correlation matrix of the vector are respectively plotted on Figure~\ref{fig:hist} and Figure~\ref{fig:corr} in Appendix~\ref{app:stat-dd}. 

\begin{rk}
    Since cumulative damages are gathered per day rather than per cycle, Miner's rule naturally yields lifetimes expressed in time units rather than in number of cycles. All lifetimes in our experimental results will therefore be expressed in days or years.
\end{rk}

\section{Numerical results}\label{s:curves}

\subsection{Survival probability curves}

In this section, we plot the survival probability for each monitored zone, as well as for the whole structure, as a function of time. Assuming a Weibull model for the random NCFs and denoting by $T^k$ and $T$ the respective random lifetime, expressed in days, of the $k$-th zone and of the structure, the identities~\eqref{eq:PNk-weibull} and~\eqref{eq:surv-struct-P} yield the formulas
\begin{equation}\label{eq:PTk}
    \forall k \in \{1, \ldots, K\}, \qquad \Pr^*(T^k > t) = \Exp^*\left[\exp\left(\log(1-p) \, \left(\sum_{i=1}^t D^k_{p,i}\right)^{m_\mathrm{W}} \right)\right]
\end{equation}
and
\begin{equation}\label{eq:PT}
    \Pr^*(T > t) = \Exp^*\left[\exp\left(\log(1-p) \, \sum_{k=1}^K\left(\sum_{i=1}^t D^k_{p,i}\right)^{m_\mathrm{W}}\right)\right],
\end{equation}
where $(D^1_{p,i}, \ldots, D^K_{p,i})_{i \geq 1}$ are independent realisations of a random vector $(D^1_p, \ldots, D^K_p)$ whose coordinates denote the daily cumulative damages at each zone $k$. Based on the sample of daily cumulative damages $\{(d^k_{p,i})_{1 \leq k \leq K}, \ 1 \leq i \leq N\}$ defined by~\eqref{eq:dkpi}, two numerical approximations may be employed to estimate the expectations in the right-hand sides of~\eqref{eq:PTk} and~\eqref{eq:PT}:
\begin{itemize}
    \item \emph{Monte Carlo approximation.} For a given number $M$ of Monte Carlo copies, 
    \begin{align}
        \Pr^*(T^k > t) &\simeq \frac{1}{M} \sum_{m=1}^M \exp\left(\log(1-p) \, \left(\sum_{i=1}^t D^{k,(m)}_{p,i}\right)^{m_\mathrm{W}} \right),
        \nonumber
        \\
        \Pr^*(T > t) &\simeq \frac{1}{M} \sum_{m=1}^M \exp\left(\log(1-p) \, \sum_{k=1}^K \left(\sum_{i=1}^t D^{k,(m)}_{p,i}\right)^{m_\mathrm{W}} \right),
        \label{eq:MC_struc}
    \end{align}
    where the vectors $(D^{k,(m)}_{p,i})_{1 \leq k \leq K}$, for $1 \leq i \leq t$ and $1 \leq m \leq M$, are drawn randomly among the sample $\{(d^k_{p,i})_{1 \leq k \leq K}, \ 1 \leq i \leq N\}$, with replacement.
    \item \emph{Deterministic equivalent damage approximation.} In accordance with~\eqref{eq:lln} and~\eqref{eq:Dkeq}, defining the deterministic equivalent daily damage for the $k$-th zone by $D^k_{p,\mathrm{eq}} = \Exp^*[D^k_p]$ yields the approximation
    \begin{align}
        \Pr^*(T^k > t) &\simeq \exp\left(\log(1-p) \, \left(t \, D^k_{p,\mathrm{eq}}\right)^{m_\mathrm{W}}\right),
        \label{eq:det_zone_k}
        \\
        \Pr^*(T > t) &\simeq \exp\left(\log(1-p) \, \sum_{k=1}^K \left(t \, D^k_{p,\mathrm{eq}}\right)^{m_\mathrm{W}}\right).
        \label{eq:det_struc}
    \end{align}
    In the right-hand side of these formulas, the deterministic equivalent daily damages may easily be estimated from the sample $\{(d^k_{p,i})_{1 \leq k \leq K}, \ 1 \leq i \leq N\}$ by the empirical mean
    \begin{equation*}
        D^k_{p,\mathrm{eq}} \simeq \frac{1}{N}\sum_{i=1}^N d^k_{p,i}.
    \end{equation*}
\end{itemize}

\begin{figure}[htpb]
    \centering
    \includegraphics[width=.8\textwidth]{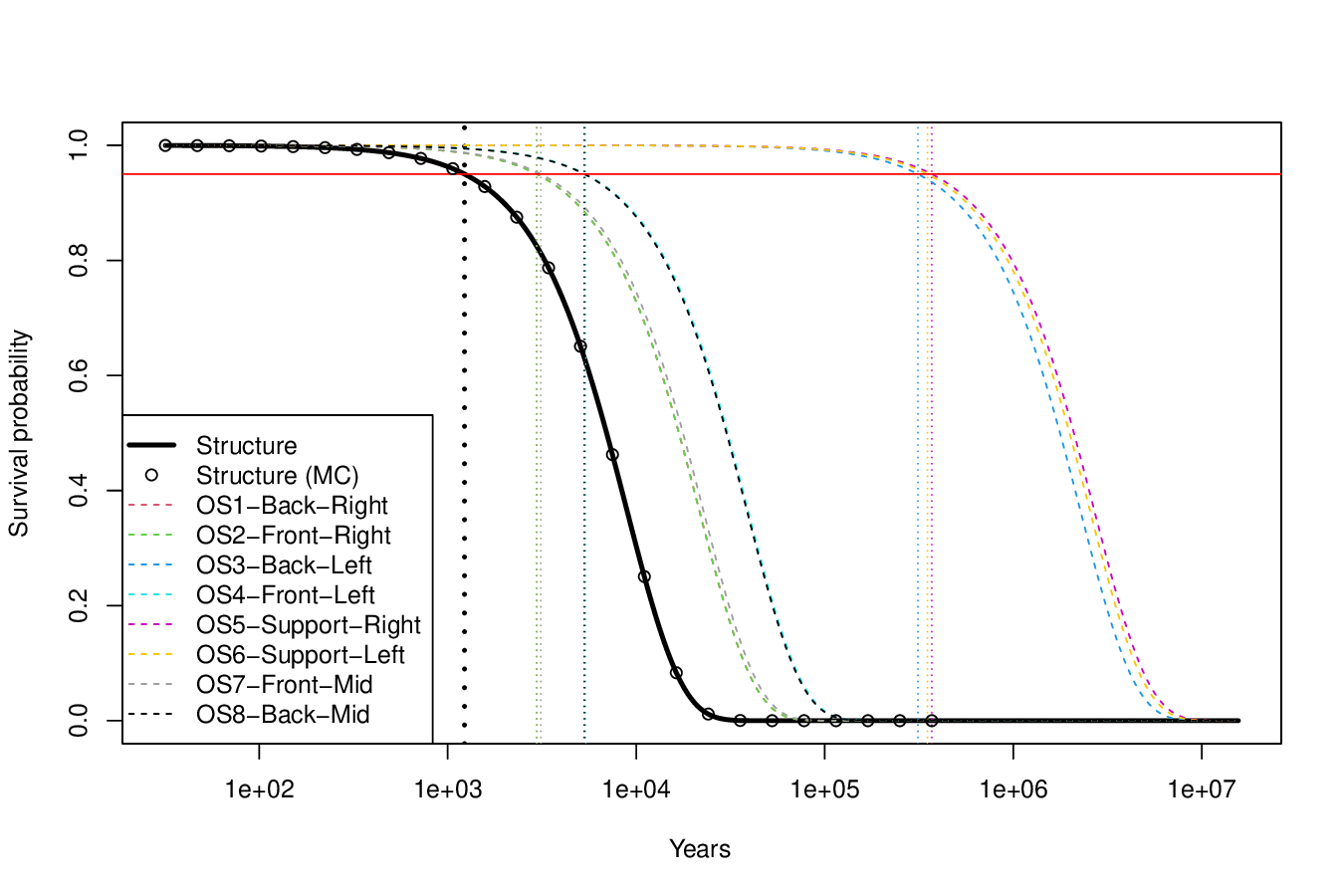}
    \caption{Survival probability of each monitored zone (in colour) and of the structure (in black), as a function of time (for $C_\mathrm{s}=1$ and $\Delta \sigma_C = 36\, \mathrm{MPa}$). The curves are computed with the deterministic equivalent damage approximation, and the Monte Carlo approximation for the survival probability of the structure is displayed with dots. The quantiles of order $p=0.05$ of the lifetimes of the zones and of the structure are represented by the vertical dotted lines. The curves for the zones 1, 2 and 7 are essentially lying one on top of each other (and likewise for the curves of the zones 4 and 8, and also for the curves of the zones 3, 5 and 6).}
    \label{fig:survival}
\end{figure}

The survival probability curves are plotted in Figure~\ref{fig:survival} for the values $C_\mathrm{s}=1$ and $\Delta \sigma_C = 36\, \mathrm{MPa}$, and $m_\mathrm{W}=1.5$. The survival probabilities for each zone are computed by the deterministic equivalent damage approximation~\eqref{eq:det_zone_k}. For the whole structure, both the deterministic equivalent damage approximation~\eqref{eq:det_struc} and the Monte Carlo approximation~\eqref{eq:MC_struc} (with $M=100$) have been considered. The agreement between both methods is excellent, which validates the use of the computationally much more efficient deterministic equivalent damage approximation.

\subsection{Miner's NCF of the structure}

In this section, we compute Miner's NCF (converted in days) of each monitored zone, obtained by a linear extrapolation of the daily cumulative damages from the database according to the standard formula
\begin{equation*}
    T^{*,k}_p = \frac{1}{D^k_{p,\mathrm{eq}}}.
\end{equation*}
We then compare the estimation of the structure's lifetime obtained by taking the minimum of Miner's NCFs of each elementary volume and the quantile of order $p$ of the structure's true, random lifetime given by the identity~\eqref{eq:barN*}. These computations are performed for $4$ scenarios, with $C_\mathrm{s}=1$ or $C_\mathrm{s}=3$, and $\Delta \sigma_C=36\, \mathrm{MPa}$ or $\Delta \sigma_C=80\, \mathrm{MPa}$.

The four scenarios are chosen as realistic ranges for the values of $C_\mathrm{s}$ and $\Delta \sigma_C$. The choice $C_\mathrm{s}=1$ with $\Delta \sigma_C=80\, \mathrm{MPa}$ is the most optimistic scenario, where we consider no additional safety factor $C_\mathrm{s}$ and a detail category $\Delta \sigma_C$ for the resistance to fatigue which is average (normative values range from $36\, \mathrm{MPa}$ to $160\, \mathrm{MPa}$). The choice $C_\mathrm{s}=3$ with $\Delta \sigma_C=36\, \mathrm{MPa}$ is the most conservative scenario, where additional stress concentration is arbitrarily modeled by a safety factor $C_\mathrm{s}=3$ and the detail category $\Delta \sigma_C$ is the worst possible, corresponding to a poor condition of the steel elements.

The results are reported in Table~\ref{tab:NCF}. For all of the four scenarios, the overestimation of the quantile of the structure's lifetime by taking the minimum quantile of the lifetimes of the monitored zones exceeds a multiplicative factor $2$. Our approach based on~\eqref{eq:det_struc} thus yields significantly more accurate results while still being very efficient from a computational viewpoint.

\begin{table}[htpb]
    \centering
    \begin{tabular}{rcccc}
        $\Delta \sigma_C$ & $36\, \mathrm{MPa}$ & $80\, \mathrm{MPa}$ & $36\, \mathrm{MPa}$ & $80\, \mathrm{MPa}$\\
        $C_\mathrm{s}$ & $3$ & $3$ & $1$ & $1$\\
        \hline
        \texttt{OS1-Back-Right} & $32$ & $580$ & $3.0\times 10^3$ & $7.3\times 10^5$\\
        \texttt{OS2-Front-Right} & $32$ & $580$ & $3.0\times 10^3$ & $7.3\times 10^5$ \\
        \texttt{OS3-Back-Left} & $550$ & $20000$ & $3.1\times 10^5$ & $6.0\times 10^7$ \\
        \texttt{OS4-Front-Left} & $61$ & $1000$ & $5.4\times 10^3$ & $1.0\times 10^6$\\
        \texttt{OS5-Support-Right} & $2800$ & $73000$ & $3.7\times 10^5$ & $1.0\times 10^8$\\
        \texttt{OS6-Support-Left} & $880$ & $26000$ & $3.5\times 10^5$ & $9.3\times 10^7$\\
        \texttt{OS7-Front-Mid} & $35$ & $600$ & $3.2\times 10^3$ & $2.4\times 10^6$\\
        \texttt{OS8-Back-Mid} & $68$ & $1000$ & $5.3\times 10^3$ & $2.0\times 10^6$\\
        \hline
        Minimal NCF & $32$ & $580$ & $3.0\times 10^3$ & $7.3\times 10^5$\\
        Lifetime quantile~\eqref{eq:barN*} & $14$ & $240$ & $1.2\times 10^3$ & $3.5\times 10^5$ \\ \\
    \end{tabular}
    \caption{Lifetime (expressed in years) of the monitored zones and (for the last two lines) of the structure, for $4$ scenarios depending on the value of the detail category $\Delta \sigma_C$ and the safety factor $C_\mathrm{s}$.}
    \label{tab:NCF}
\end{table}

\section{Conclusion}\label{s:conclusion}

In this article, the probabilistic methodology introduced in~\cite{Car23} to compute the survival probability of a structure subjected to cyclic loading with random amplitude is put into practice on an industrial structure for which experimental data are available. We provide a comprehensive comparison of the results provided by this methodology with those obtained using the standard engineering approach based on the combination of Miner's cumulative damage rule and the weakest link principle among the monitored zones.

In principle, the survival probability of the structure given by the formula~\eqref{eq:surv-struct-P*} writes as an expectation with respect to the law of the sequence of daily cumulative damages measured by each sensor, and therefore its numerical evaluation may require a computationally costly procedure based on the Monte Carlo method. Our first finding is that replacing the stochastic sequence of daily damages by its expectation (a quantity that we called the \emph{equivalent deterministic daily damage}) makes the numerical evaluation of the survival probability~\eqref{eq:surv-struct-P*} straightforward (see~\eqref{eq:surv-struct-P*-eq}) while preserving the quality of its estimation.

Once this equivalent deterministic damage is estimated, the quantile of order $p$ of the structure's fatigue life can be assessed according to the following methodology: first, Miner's rule is applied at the level of each monitored zone to estimate its individual lifetime $T^{*,k}_p$; second, the structure's lifetime quantile is obtained by the formula~\eqref{eq:barN*}. This formula systematically yields an estimated lifetime which is smaller (by at least a factor 2 in our numerical examples) than taking the minimum of the zones' lifetimes $T^{*,k}_p$. This shows that our probabilistic procedure significantly reduces the risk of overestimating the fatigue life of the structure.


\appendix

\section{Statistics of daily cumulative damages}\label{app:stat-dd}

\begin{figure}[htpb]
    \centering
    \includegraphics[width=.45\textwidth]{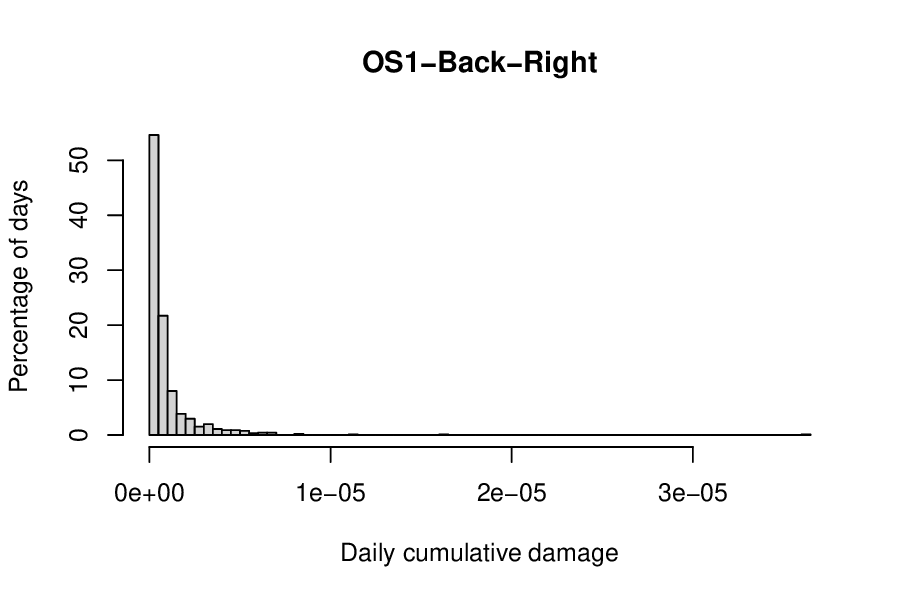}\includegraphics[width=.45\textwidth]{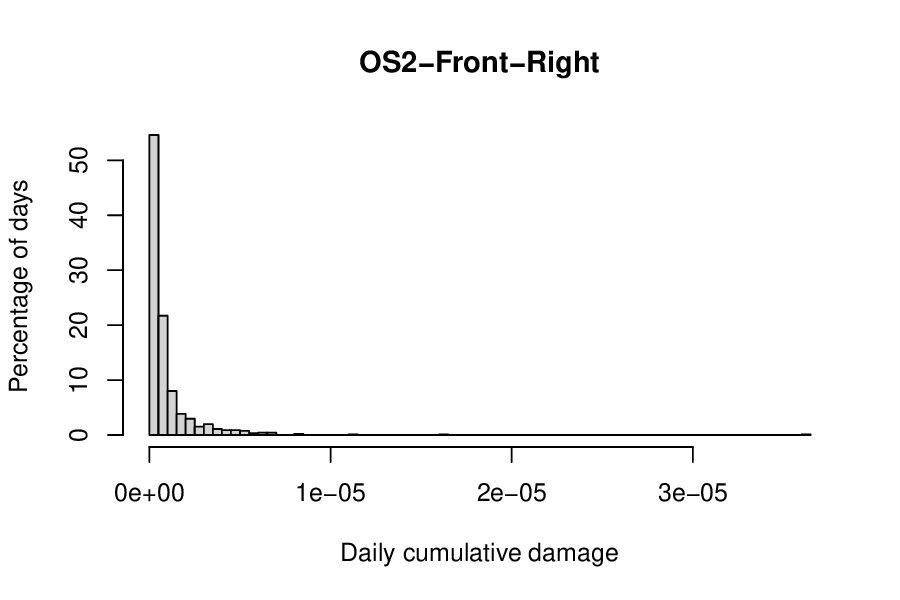}\\
    \includegraphics[width=.45\textwidth]{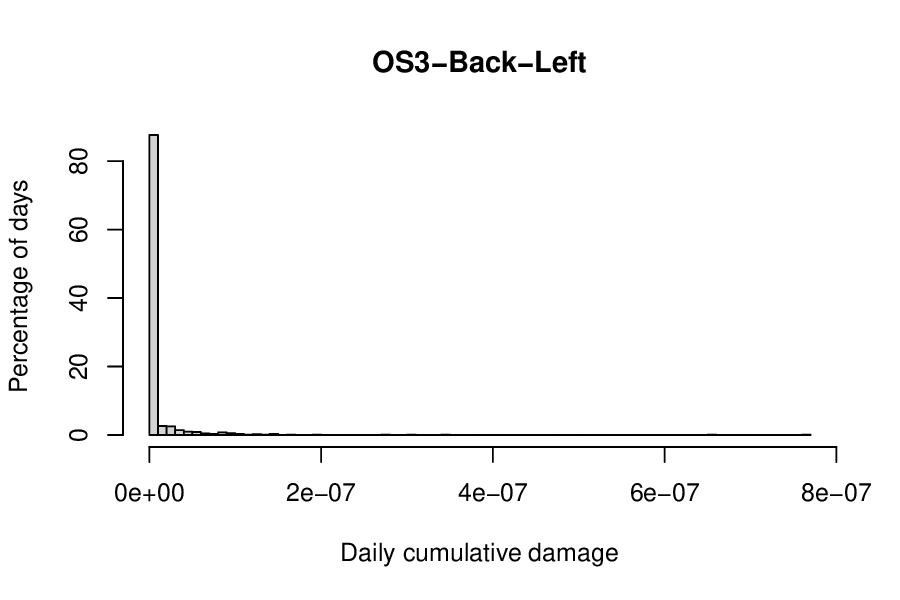}\includegraphics[width=.45\textwidth]{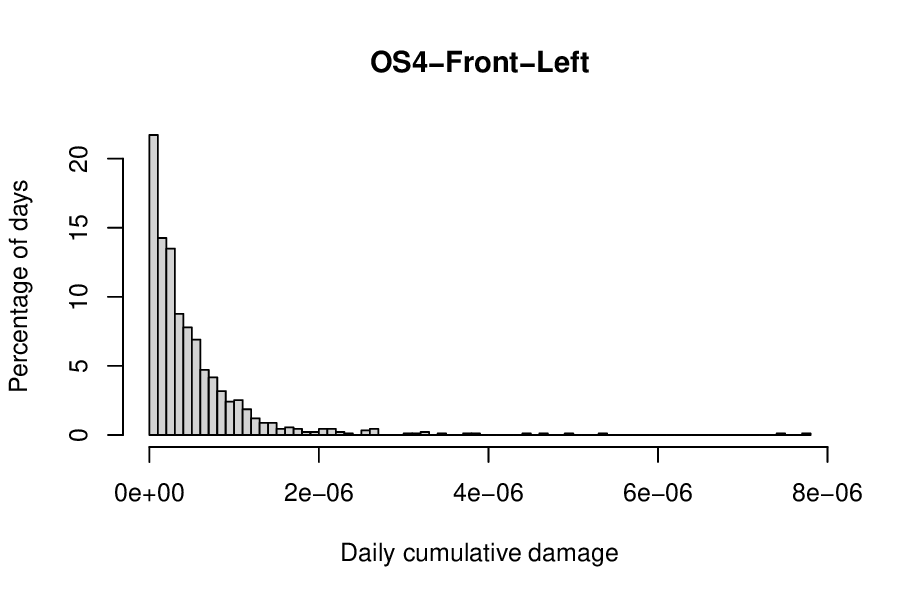}\\
    \includegraphics[width=.45\textwidth]{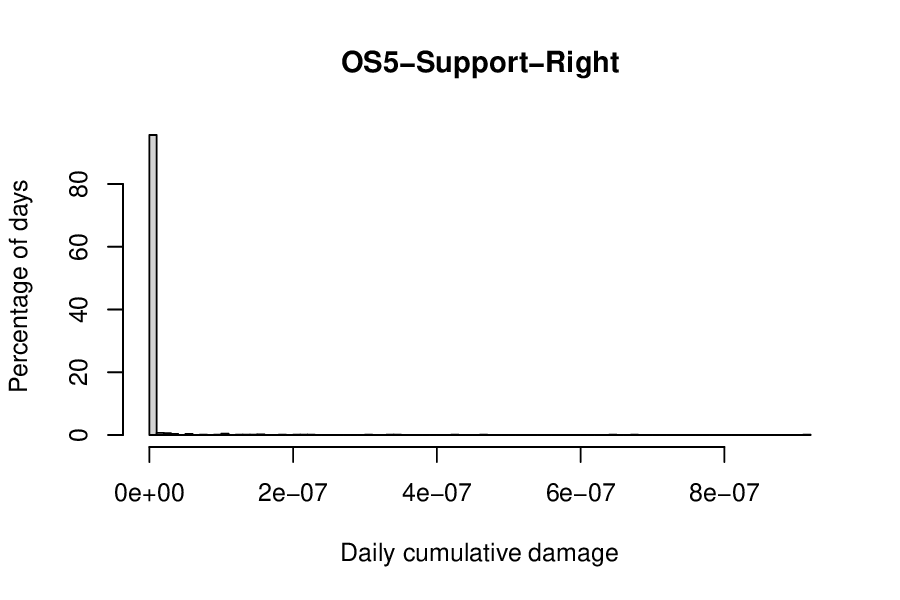}\includegraphics[width=.45\textwidth]{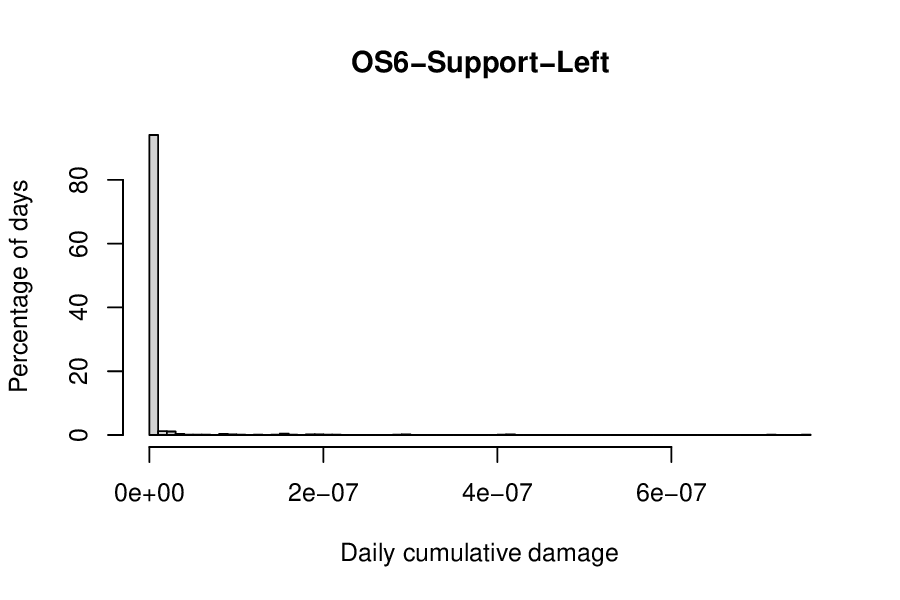}\\
    \includegraphics[width=.45\textwidth]{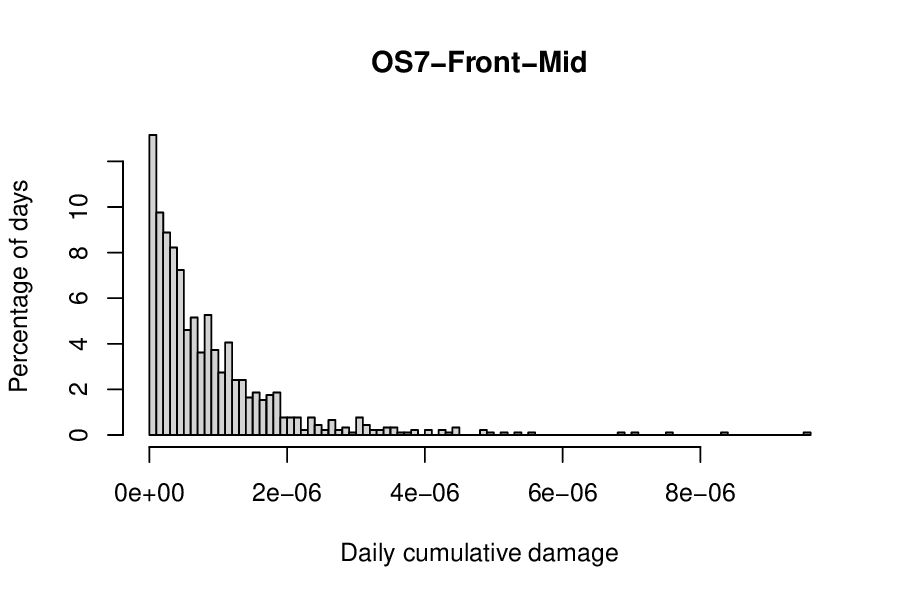}\includegraphics[width=.45\textwidth]{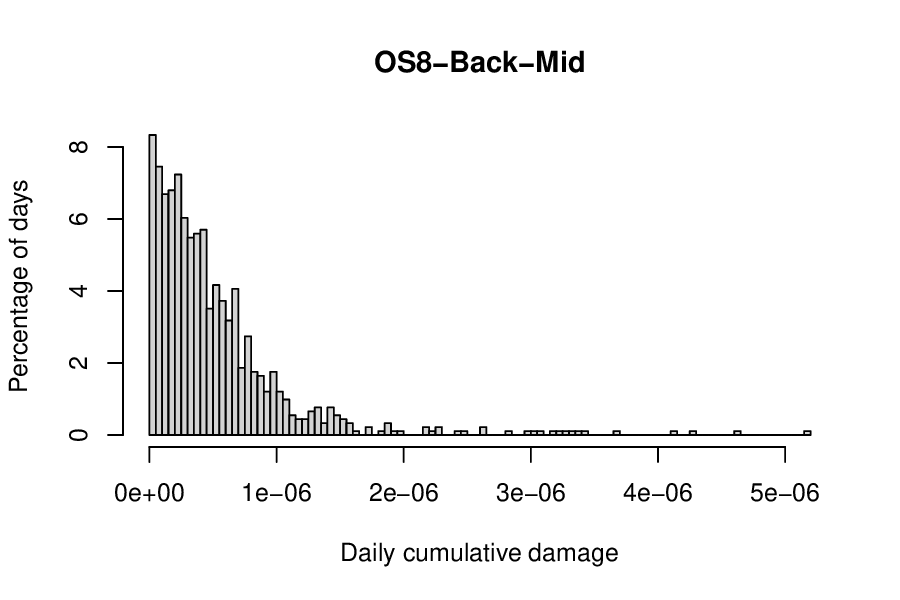}
    \caption{Histograms of the daily cumulative damages at each sensor. The $y$-axis provides the percentage of days (over the monitoring period) for which the observed cumulative damage lies into the corresponding class. For sensors \texttt{OS3-Back-Left}, \texttt{OS5-Support-Right} and \texttt{OS6-Support-Left}, the percentage of days with no cumulative damage is $84\%$, $94\%$ and $92\%$, respectively. On these days, either no cycle has been recorded at all, or all associated severities are below the cut-off limit of the S-N curve (see Figure~\ref{fig:sn}).}
    \label{fig:hist}
\end{figure}

\begin{figure}[htpb]
    \centering
    \includegraphics[width=.6\textwidth]{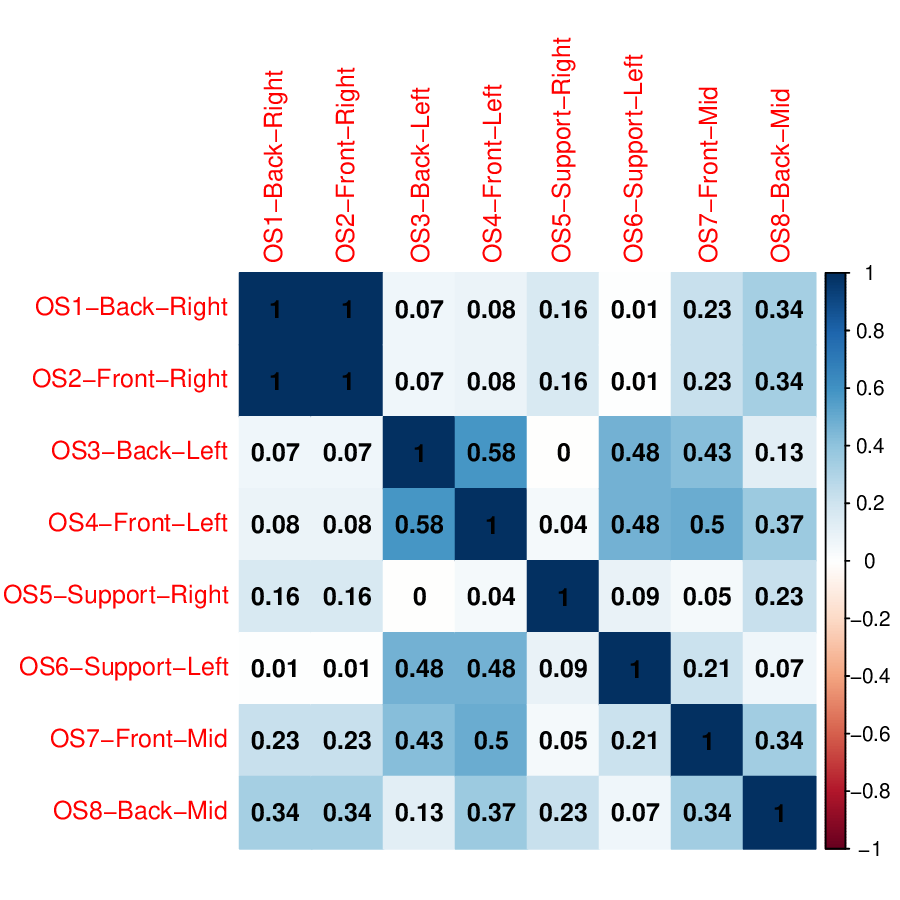}
    \caption{Correlations between the daily cumulative damages recorded at each sensor.}
    \label{fig:corr}
\end{figure}

In this Appendix, we describe the statistics of the vector of daily cumulative damages $(D^k_p)_{1 \leq k \leq K}$ where, we recall, $k$ is the zone (i.e. elementary volume) index. These results have been computed from the experimental database (yielding the values of $\eps$ and thus of $S^k$, see~\eqref{eq:S-eps} where we have chosen $C_\mathrm{s}^k = C_\mathrm{s} = 1$) and using~\eqref{eq:dkpi} where we have chosen $\Delta \sigma^k_C = \Delta \sigma_C = 36\, \mathrm{MPa}$. The marginal histograms of each coordinate and the correlation matrix of the vector are plotted on Figure~\ref{fig:hist} and Figure~\ref{fig:corr}, respectively.

\section*{Acknowledgements}

We would like to thank Alain Ehrlacher for many stimulating and enlightening discussions about the work reported in this article. This research received the support of OSMOS Group, France, as part of its effort to develop new solutions for the Structural Health Monitoring of civil and industrial assets.

\bibliographystyle{plain}
\bibliography{biblio-miner}

\end{document}